\documentclass[onecolumn,prc,preprintnumbers,nofootinbib,superscriptaddress,showpacs,a4]{revtex4-1}
\usepackage{epsfig,graphicx,float,pst-all}
\usepackage{mathrsfs}
\usepackage[utf8]{inputenc}
\usepackage{rotating}
\usepackage{url}
\usepackage{esvect}
\usepackage{subfigure}
\usepackage{multirow}
\usepackage{amsmath}
\usepackage{amssymb}
\usepackage{xcolor}
\usepackage{natbib}
\usepackage{hyperref}
\usepackage[warn]{textcomp}
\usepackage{gensymb}
\usepackage{ulem}
\usepackage{footnote}
\usepackage{siunitx}
\usepackage{comment}
\usepackage{bm}
\usepackage{orcidlink}
\usepackage{outlines}

\newcommand{\nucl}[2]{^{#1}\mathrm{#2}}

\begin{document}

\title{Is  $\nucl{40}{Mg}$ a Borromean halo nucleus? \\A case built on the electric-dipole response\footnote{Invited contribution at the \textbf{XXXVIII Mazurian Lakes Conference on Physics}, Piaski, Poland, August 31 - September 6, 2025.}}
\author{Jagjit Singh\orcidlink{0000-0002-3198-4829}}
\email{Jagjit.Singh@manchester.ac.uk}
\affiliation {Department of Physics and Astronomy, The University of Manchester, Manchester M13 9PL, UK}
\affiliation{Department of Physics, Akal University, Talwandi Sabo, Bathinda, Punjab 151302, India}
\affiliation{Research Center for Nuclear Physics (RCNP), Osaka University, Ibaraki 567-0047, Japan}
\author{J. Casal\orcidlink{0000-0002-5997-5860}}
\email{jcasal@us.es}
\affiliation{Departamento de F\'{i}sica At\'{o}mica, Molecular y Nuclear, Facultad de F\'{i}sica, Universidad de Sevilla, Apartado 1065, E-41080 Sevilla, Spain}
\author{N. R. Walet\orcidlink{0000-0002-2061-5534}}
\affiliation{Department of Physics and Astronomy, The University of Manchester, Manchester M13 9PL, UK}
\author{W. Horiuchi\orcidlink{0000-0003-0039-3783}}
\affiliation{Department of Physics, Osaka Metropolitan University, Osaka 558-8585, Japan}
\affiliation{Nambu Yoichiro Institute of Theoretical and Experimental Physics (NITEP), Osaka Metropolitan University, Osaka 558-8585, Japan}
\affiliation{RIKEN Nishina Center, Wako 351-0198, Japan}
\affiliation{Department of Physics, Hokkaido University, Sapporo 060-0810, Japan}
\author{W. Satu{\l}a\orcidlink{0000-0003-0203-3773}}
\affiliation{Institute of Theoretical Physics, Faculty of Physics, University of Warsaw, ul. Pasteura 5, PL-02-093 Warsaw, Poland}
\date{\today}

\begin{abstract}
We investigate the low-energy electric-dipole response of $\nucl{40}{Mg}$ using a $\nucl{38}{Mg}+n+n$ three-body model. This model is implemented using a three-body hyperspherical formalism with an analytical transformed harmonic oscillator basis. In this study, two different neutron-neutron interactions are considered: a scalar Gaussian density-dependent central potential and a more realistic finite-range potential which includes central, spin-orbit, and tensor components. We examine how electric-dipole response is affected by the choice of the interaction.
\end{abstract}
\maketitle

\section{Introduction}
The year $2025$ marks the $40^\text{th}$ anniversary of the introduction of the concept of a neutron halo \cite{Halo40_2025}, a landmark in the exploration of exotic nuclear structure near the neutron dripline. The idea originates from a surprising experimental observation made in $1985$, when unusually large interaction cross sections were measured in light neutron-rich nuclei such as $\nucl{11}{Li}$ \cite{Tanihata1985}. These  results hinted that some neutrons were located far outside the normal radius of the nucleus. By 1987, this unexpected type of structure had a name: the neutron halo \cite{Hansen1987}. Since then, halo nuclei have remained a central topic in experimental nuclear physics, where two robust experimental signatures are used to identify such nuclei: Anomalously large matter radii extracted from interaction cross-section measurements that point to a diffuse matter distribution, and enhanced low-lying electric dipole strength observed via Coulomb dissociation as evidence for their weak binding and cluster-like dynamics.

The rapid development of Rare Isotope Beam (RIB) facilities has opened experimental access to neutron-rich areas of the nuclear chart that were previously beyond reach. This has led to detailed studies of the neutron-rich frontier, especially near the traditional shell closures at $N=20$ \cite{Bagchi2020, Revel2020,Singh2020,Fortunato2020,Casal2020,GSingh2022} and $N=28$ \cite{Ahn2022,Zhang2023,SinghPLB2024,Singh2025}. In these areas of the nuclear chart, the combination of weak binding, intruder configurations, and evolving shell structure creates the conditions for the emergence of halos \cite{Li2024}. Indeed, the recent identification of $\nucl{29}{F}$ \cite{Bagchi2020,Casal2020} as a two-neutron Borromean halo nucleus extends the phenomenon beyond the light mass region. This demonstrates that halo formation can persist even in nuclei with substantial deformation and strong configuration mixing, provided that the two-neutron separation energy remains small and the last neutron occupies a low-angular-momentum orbital.

These developments raise the question of where the next halo systems will emerge. The region around the $N=28$ shell closure is a particularly promising candidate due to the known weakening of the shell gap for small proton numbers \cite{Zhang2023,SinghPLB2024,Singh2025}. Our recent three-body (core$+n+n$) calculations for $\nucl{39}{Na}$ and $\nucl{40}{Mg}$, combined with Glauber reaction model analyses, predicts the possibility of enhanced matter radii and reaction cross sections, suggesting halo structures near this weakened shell closure \cite{SinghPLB2024,Singh2025}.

If there is a halo in $\nucl{40}{Mg}$, it is expected to have important consequences for direct reactions. In particular, the halo should be evident in the  electric-dipole ($E1$) response, which can be probed via high-energy Coulomb dissociation. Its presence also affects low-energy elastic scattering and breakup processes. In the present work, we focus on the electric-dipole response of $\nucl{40}{Mg}$, employing an effective three-body description in which the nucleus is modeled as a $\nucl{38}{Mg}$ core coupled to two valence neutrons. This framework allows us to probe the evolution of the dipole strength distribution as a function of the two-neutron separation energy, which is uncertain since it has not been measured yet, and to quantify the degree of halo formation. In analogy with the case of light halo nuclei such as $\nucl{6}{He}$ or $\nucl{11}{Li}$, strong low-energy dipole transitions would provide a direct signature of the extended spatial distribution of the two valence neutrons.

\section{Theoretical Framework}
In principle, the wavefunction is a function of the coordinates $\boldsymbol{r}_x$,  the relative distance between the two neutrons and $\boldsymbol{r}_y$, the distance between the core and the center of mass of neutrons. However, we describe the three-body $\text{core}+n+n$ system using a hyperspherical formalism, for more details see Refs.~\cite{ZHUKOV1993,JCasal13}. 
The wave function of total angular momentum $j$ is expressed in the scaled coordinates $\boldsymbol{x}=\boldsymbol{r}_{x}\sqrt{\frac{1}{2}}$ and $\boldsymbol{y}=\boldsymbol{r}_{y}\sqrt{\frac{2A}{A+2}}$, where $A$ is the core mass number, as
\begin{equation}
  \psi^{j\mu}(\rho,\Omega) = \rho^{-5/2}\sum_{\beta}\chi_{\beta}^{j}(\rho)\mathcal{Y}_{\beta}^{j\mu}(\Omega)\,.
  \label{eq:3bwf}
\end{equation}
Here  $\rho=\sqrt{x^2+y^2}$ is the hyper-radius and the hyper-angular coordinates $\Omega\equiv\{\alpha,\hat{x},\hat{y}\}$, with  $\alpha=\arctan{(x/y)}$.  
The channel index $\beta=\{K,l_x,l_y,l,S_x,J\}$ labels the coupling order of the hyperspherical harmonics and spin functions in $\mathcal{Y}_\beta^{j\mu}(\Omega)$, were $l_x$ and $l_y$ couple to $l$, and $l$ and $S_x$ to angular momentum $j$. The hyperspherical quantum number $K$ satisfies the condition that $(K-l_x-l_y)/2$ must be a non-negative integer. 
If we assume the core is spherical and has angular momentum zero, $J=j$, and the number of channels reduces accordingly. The expansion is truncated at $K\leq K_{\max}$. For a more complete definition of the notation used  the reader should consult Refs.~\cite{Casal2020,Casal19B2020}.

The radial functions $\chi_\beta^j(\rho)$ are obtained by solving the coupled hyperradial equations with pairwise potentials $V_{ij}$, requiring the coupling matrix
\begin{equation}
V_{\beta'\beta}^{j\mu}(\rho)=\left\langle \mathcal{Y}_{\beta }^{j\mu}(\Omega)\Big|V_{12}+V_{13}+V_{23} \Big|\mathcal{Y}_{\beta'}^{ j\mu}(\Omega) \right\rangle_\Omega.
\label{eq:3bcoup}
\end{equation}
In our case, $V_{12}=V_{nn}$ and $V_{13}=V_{23}=V_{\text{core-}n}$. In addition to the binary interactions, it is customary to include a phenomenological three-body force in Eq.~\ref{eq:3bcoup} to account for effects beyond the strict three-body framework. This force is typically modeled as a hyper-scalar Gaussian potential, $V_{3b}(\rho)=v_{3b}e^{-{\left(\rho/\rho_o\right)}^2}$, where $\rho_0\approx5$–6 fm and $v_{3b}$ is chosen to reproduce the ground-state energy. As an alternative, we can also use a density-dependent modification of the central $nn$ interaction \cite{Hagino05,Casal19B2020}, 
\begin{equation}
    \tilde{V}_{nn}^{\rm c}(r_x,r_y)=V_{nn}^{\rm c}(r_x)\left(1+\frac{v_0}{1+\exp{[(r_{y}-R_0)/a_0]}}\right),
    \label{eq:nndensdep}
\end{equation}
with $a_0=0.67$ fm, $R_0=1.27A^{1/3}$, and $v_0$ is once again chosen to reproduce the ground-state energy. This density-dependent interaction vanishes at large core–dineutron distances and shifts the ground-state energy through its short-range behavior.
Rather than solving the coupled hyperradial equations directly, we expand 
the radial functions $\chi_\beta^{j}(\rho)$ in a finite basis and diagonalize the Hamiltonian, obtaining bound states and continuum pseudo states (PS) for discrete eigenvalues $\varepsilon_n$. 
Here, we use the Analytically Transformed Harmonic Oscillator basis, which allows additional control over the radial extent of the wave functions, and also low-energy PS density.
The dipole strength for $\text{ground state}\rightarrow j$ continuum transitions is $B(E1) = |\langle \textit{gs} ||\widehat{O}_{E1}||\varepsilon_n,j\rangle|^2$ with 
\begin{equation}
    \widehat{O}_{E1} = Ze\sqrt{\frac{2}{A(A+2)}}y Y_{1 M}(\widehat{y}).
    \label{eq:op}
\end{equation}
The pseudo-state energies are of course discrete, and their dipole strength is  folded with a Poisson kernel to obtain a smooth distribution while preserving the integrated total strength.

\section{Results and Discussions}
Following the methodology outlined above, we now turn our attention to nucleus $\nucl{40}{Mg}$, which has been proposed as a potential two-neutron halo candidate, based on various theoretical predictions~\cite{Caurier2014,Nakada2018,Macchiavelli2022}. The two-neutron separation energy $(S_{2n})$ of $\nucl{40}{Mg}$ is not experimentally measured but we can use the evaluated (i.e., estimated based on systematics) value $S_{2n} = 0.670 \pm 0.710$ MeV~\cite{Wang2021}. 
In this work, we focus on the electric dipole response; the corresponding ground-state properties have already been investigated in our recent work Ref.~\cite{SinghPLB2024}.

To model the three-body system, we adopt a simplified approach and assign an inert angular momentum zero state to $^{38}$Mg. A more refined treatment would require detailed spectroscopic data for $\nucl{39}{Mg}$ and more sophisticated $\nucl{38}{Mg}+n$ interactions to account for possible core excitations and/or level splitting. Given the absence of experimental data, we describe the $\nucl{38}{Mg}+n$ interaction using a simple Woods-Saxon potential with central and spin-orbit components,
\begin{equation}
    V_{{\rm core}+n}(r) = \left(-V^{(l)}_0+V_{ls}\lambda_\pi^2 \vec{l}\cdot\vec{s}\frac{1}{r}\frac{d}{dr}\right)\frac{1}{1+\exp[(r-R_c)/a]}\,.\label{eq:WS}
\end{equation} 
Here, $V^{(l)}_0$ is $l$-dependent, and $R_c = r_0 A_c^{1/3}$ with $A_c$ being the mass number of the core. The spin-orbit term is expressed using the pion Compton wavelength $\lambda_\pi = 1.414$ fm. Following the systematics of Ref.~\cite{HOR10}, we adopt $V_{ls} = 16.842$ MeV for the $\nucl{38}{Mg}+n$ interaction. The radius parameter $r_0 = 1.25$ fm is consistent with earlier studies on neutron-rich nuclei ~\cite{Singh2020,HOR10}, and we consider three interaction scenarios as detailed in upper-panel of Table 1 of Ref.~\cite{SinghPLB2024}.

For the neutron-neutron ($n$-$n$) subsystem, we employ two types of interactions. The first is a semi-realistic finite-range potential, the Gogny-Pires-Tourreil (GPT) potential~\cite{Gogny70}, which includes central, spin-orbit, and tensor components which has been successfully used in previous three-body studies~\cite{ZHUKOV1993, FACE2004, Singh2020, GSingh2022}. The second is a simplified Gaussian central potential of the form $V_{nn}(r_x) = S \exp[-(r_x/b)^2]$, with parameters $(S, b)$ chosen to reproduce the known $n$-$n$ scattering length $a_s = -15$ fm \cite{Casal19B2020}. To explore finite-range effects, we consider four parameter sets  \cite{Casal19B2020}: 
\begin{align}
   \text{ddG-1:} &(S, b) = (-675.0\,\text{MeV}, 0.4\,\text{fm}),\nonumber\\
   \text{ddG-2:} &(S, b) =(-164.0\,\text{MeV}, 0.8\,\text{fm}),\nonumber\\ 
   \text{ddG-3:} &(S, b) =(-24.22\,\text{MeV}, 2.0\,\text{fm}), \nonumber\\
   \text{ddG-4:} &(S, b) =(-8.75\,\text{MeV}, 3.2\,\text{fm})\,.\label{eq:ddGpars}
\end{align} In the limit of small $b$, this Gaussian potential, when supplemented with a density-dependent term (ddG) [see Eq.~(\ref{eq:nndensdep})], is very similar to the density-dependent contact pairing interaction used in earlier three-body models~\cite{Hagino05}. 

We generate the $1^-$ continuum states using a THO basis with parameters $b = 0.7$ fm, $\gamma = 1.0$ fm$^{1/2}$, $N = 24$, and include partial waves up to $K_{\text{max}} = 38$. The parameters $(b, \gamma)$ transform the Gaussian asymptotic behavior of the harmonic oscillator functions into an exponential decay, which improves the convergence of the calculations with respect to the number of basis functions $N$. Additionally, the ratio $\gamma/b$ controls the hyperradial range of the basis functions.  Since there are no bound $1^-$ states, ensuring convergence with respect to the choice of $K_{\text{max}}$ and $N$ is not straightforward. We assume that achieving a stable $B(E1)$ distribution is sufficient to ensure convergence, but this can be challenging for a very low-energy peak in strength. For these calculations, we use the same $v_{3b}$ (for GPT) and $v_0$ (when using ddG) in the $1^-$ continuum, that reproduce the $0^+$ ground state. Thus,  the $B(E1)$ calculations involve no additional parameter fitting.

\subsection{$B(E1)$ distribution for $\nucl{40}{Mg}$ with GPT interaction}
\begin{figure}[t]
\centering
\includegraphics[width=0.78\linewidth]{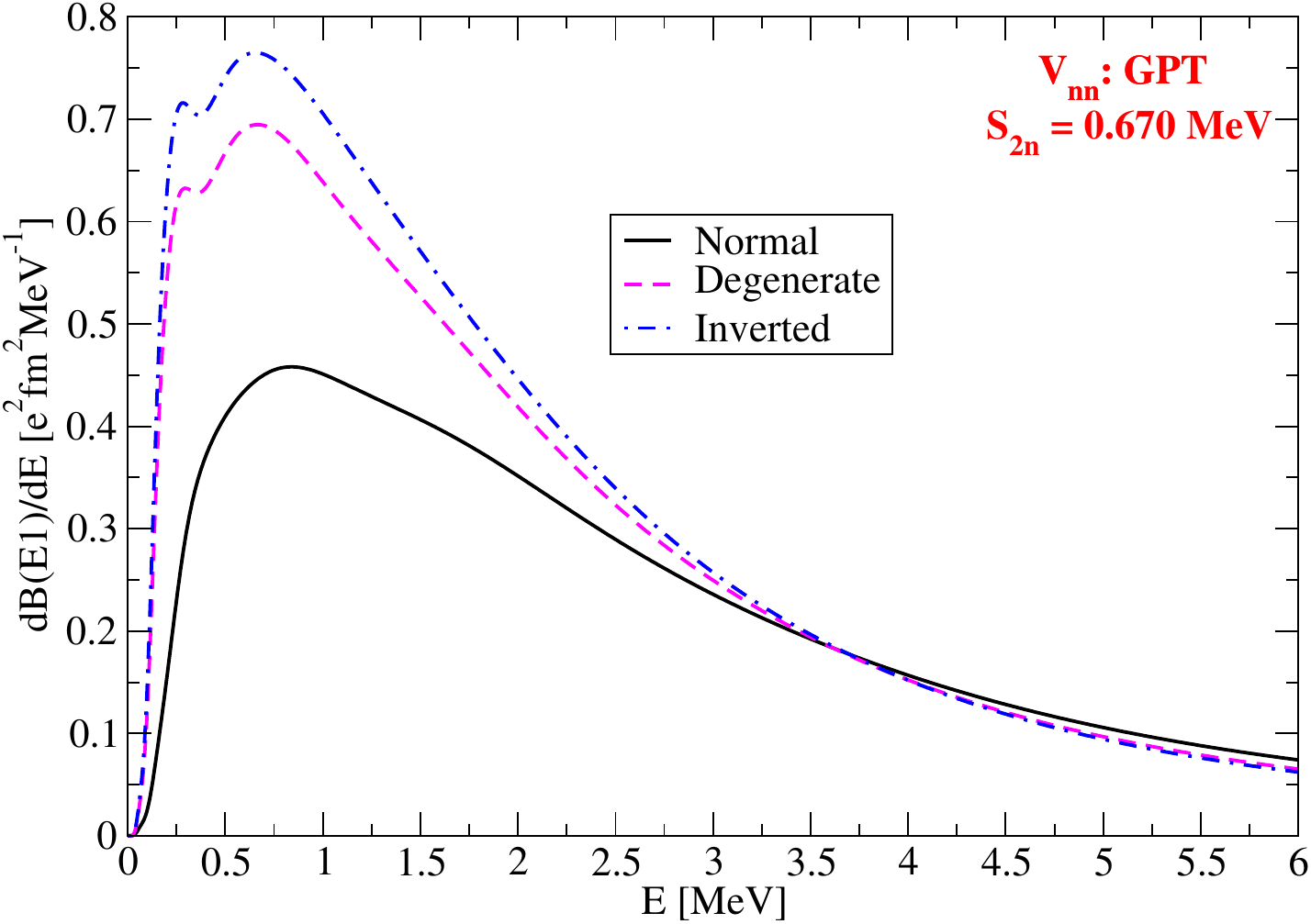}
\caption{$B(E1)$ distribution for $\nucl{40}{Mg}$ using the GPT interaction and a Gaussian three-body force for
three different $\nucl{38}{Mg}+n$ scenarios discussed in the text.}
\label{fig:BE1GPT}
\end{figure}

Figure \ref{fig:BE1GPT} shows the calculated $B(E1)$ distribution for $\nucl{40}{Mg}$ obtained with the GPT interaction and a three-body force for three different $\nucl{38}{Mg}+n$ scenarios labeled by the order of the valence orbitals $1f_{7/2}$ and $2p_{3/2}$ (described in detail in the upper panel of Table 1 in Ref.~\cite{SinghPLB2024}): Normal (highest spin lowest, solid black line), Degenerate (dashed magenta line), and Inverted (dot-dashed blue line). In every case the ground state is constrained to reproduce the central value of the evaluated two-neutron separation energy $S_{2n}=0.670$ MeV. The curves are produced by smoothing the discrete $B(E1)$ strengths with Poisson functions. We show  converged distributions up to 6 MeV energy beyond the threshold in Fig. \ref{fig:BE1GPT}. 

The integrated dipole strength up to 6 MeV is $1.44$, $1.82$, and $1.94\,e^2\mathrm{fm}^2$ for the Normal, Degenerate, and Inverted scenarios, respectively. These represent large fractions of the corresponding dipole cluster sum rule values, which are $1.88$, $2.17$, and $2.26\,e^2\mathrm{fm}^2$ respectively. Thus the dipole strength up to $6\,\mathrm{MeV}$ amounts to $76\%$, $84\%$, and $86\%$ of the sum-rule total. Indeed, extending the integration to 15 MeV brings the total strength close to the exact sum-rule limit. Such a concentration of $B(E1)$ strength at low continuum energies is a well-known signature of neutron-halo structure. The integrated dipole strength for the Inverted scenario is significantly larger than those observed in classic two-neutron halo nuclei. For example, the dipole strength of $\nucl{6}{He}$ is measured to be 1.1(1) e$^2$fm$^2$ up to 6 MeV \cite{SUN2021}, for $\nucl{11}{Li}$ it is 1.42(0.18) e$^2$fm$^2$ up to 3 MeV \cite{Nakamura2006}, and for $\nucl{19}{B}$ it reaches 1.64(0.06) e$^2$fm$^2$ up to 6 MeV \cite{Cook2020}.

As a final comment on the $B(E1)$ distribution, we note that this observable (and the corresponding cross section) is highly sensitive to the ground-state radius and configuration mixing. A marked enhancement in $B(E1)$ distribution appears when the intruder orbital $2p_{3/2}$ dives below the normal $1f_{7/2}$ orbital, i.e., the Inverted scenario. In this case the ground state shows a larger matter radius, $3.767$\,fm, compared with $3.726$\,fm for the Normal scenario and $3.758$ fm for the Degenerate case. This behavior arises from the shifting balance between $(2p_{3/2})^2$ and $(1f_{7/2})^2$ occupancies while maintaining the same $S_{2n}=0.670$ MeV. The respective percentage occupancies of $(2p_{3/2})^2$ and $(1f_{7/2})^2$ are (34.178$\%$, 53.657$\%$) for the Normal scenario, (44.731$\%$, 42.243$\%$) for the Degenerate scenario, and (50.014$\%$, 36.663$\%$) for the Inverted scenario. These values appear in the Degenerate picture of the upper left panel of Fig.~2 in Ref.~\cite{SinghPLB2024}. Because the radii differ by only about $1\%$ with $S_{2n}$ kept fixed, radius measurements alone cannot clearly distinguish these scenarios. Data from knockout or transfer reactions, which probe the ground-state wave function, would be much more effective in identifying them.

\subsection{$B(E1)$ distribution for $\nucl{40}{Mg}$ with GPT interaction for different $S_{2n}$ }
\begin{figure}[h!]
\centering
\includegraphics[width=0.78\linewidth]{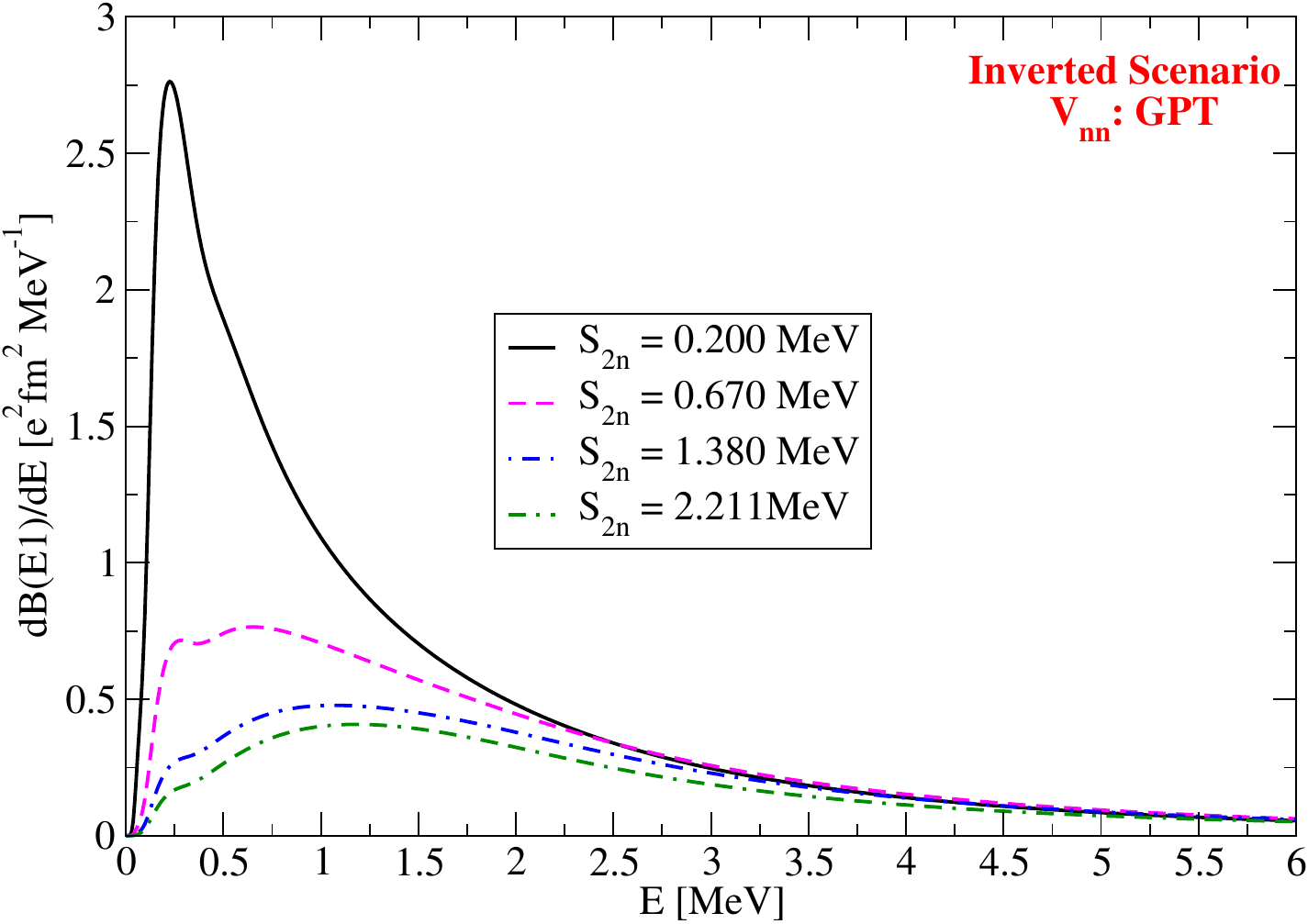}
\caption{$B(E1)$ distribution for $\nucl{40}{Mg}$ 
calculated with the GPT interaction and a three-body force in the Inverted scenario. The results are shown, as labelled,  for four choices of the two-neutron separation energy $S_{2n}$. For details, see text.}
\label{fig:BE1GPTs2n}
\end{figure} 

To further investigate the effect of the two-neutron separation energy ($S_{2n}$) on the $B(E1)$ distribution for $\nucl{40}{Mg}$, we examine the matter radii and the partial-wave composition of the ground state, focusing on the $(2p_{3/2})^2$ and $(1f_{7/2})^2$ configurations. Given the significant uncertainties in the evaluated $S_{2n}$ value for $\nucl{40}{Mg}$ \cite{Wang2021}, we explore the $B(E1)$ distribution across a range of $S_{2n}$ values by adjusting the three-body potential strength. For $\nucl{40}{Mg}$, a three-body force with $V_{3b} > 0$ is necessary to reproduce the full range of the evaluated value of $S_{2n}$: 0.200, 0.670, and 1.380 MeV. Here, we replace the rigorous lower limit of $0$, which corresponds to an unbound state, contradicting experimental evidence, which is supported by a shallow bound value. For purpose of comparison, we also consider a deeply-bound case with $V_{3b} = 0$, where $S_{2n} = 2.211$ MeV. As expected, shallower ground states exhibit enhanced $p$-wave contributions and reduced $f$-wave occupancy. In contrast, deeper binding results in degenerate and normal occupancy of these states, as illustrated in Fig.~2 of Ref.~\cite{SinghPLB2024}. 

It is evident in Fig.~\ref{fig:BE1GPTs2n} that the shallowest ground state ($S_{2n}=0.200$\,MeV), with the highest $p$-wave occupancy and largest matter radius, yields the strongest $B(E1)$ response. The integrated dipole strength up to 6 MeV is 3.10, 1.94, 1.43, and 1.18 e$^2$fm$^2$ for the $S_{2n}=$ 0.200, 0.670, 1.380, and 2.211, respectively. These values represent substantial fractions of the corresponding dipole cluster sum rule limits, amounting to 93$\%$, 86$\%$, 80$\%$, and 76$\%$ of the total for $S_{2n}$ values of 0.200, 0.670, 1.380, and 2.211 MeV, respectively. Our results indicate a pronounced halo structure in the shallowest case.

We see that modifying the ground-state energy shifts both the total $B(E1)$ strength and the position of the peak. This sensitivity suggests that the observed peak probably does not originate from a true resonance. To clarify this point, a more rigorous identification following, for instance, the method of Ref.~\cite{Casal2019} is needed and will be addressed in future work.

\subsection{$B(E1)$ distribution for $\nucl{40}{Mg}$ with ddG vs. GPT interaction}
\begin{figure}[h!]
\centering
\includegraphics[width=0.78\linewidth]{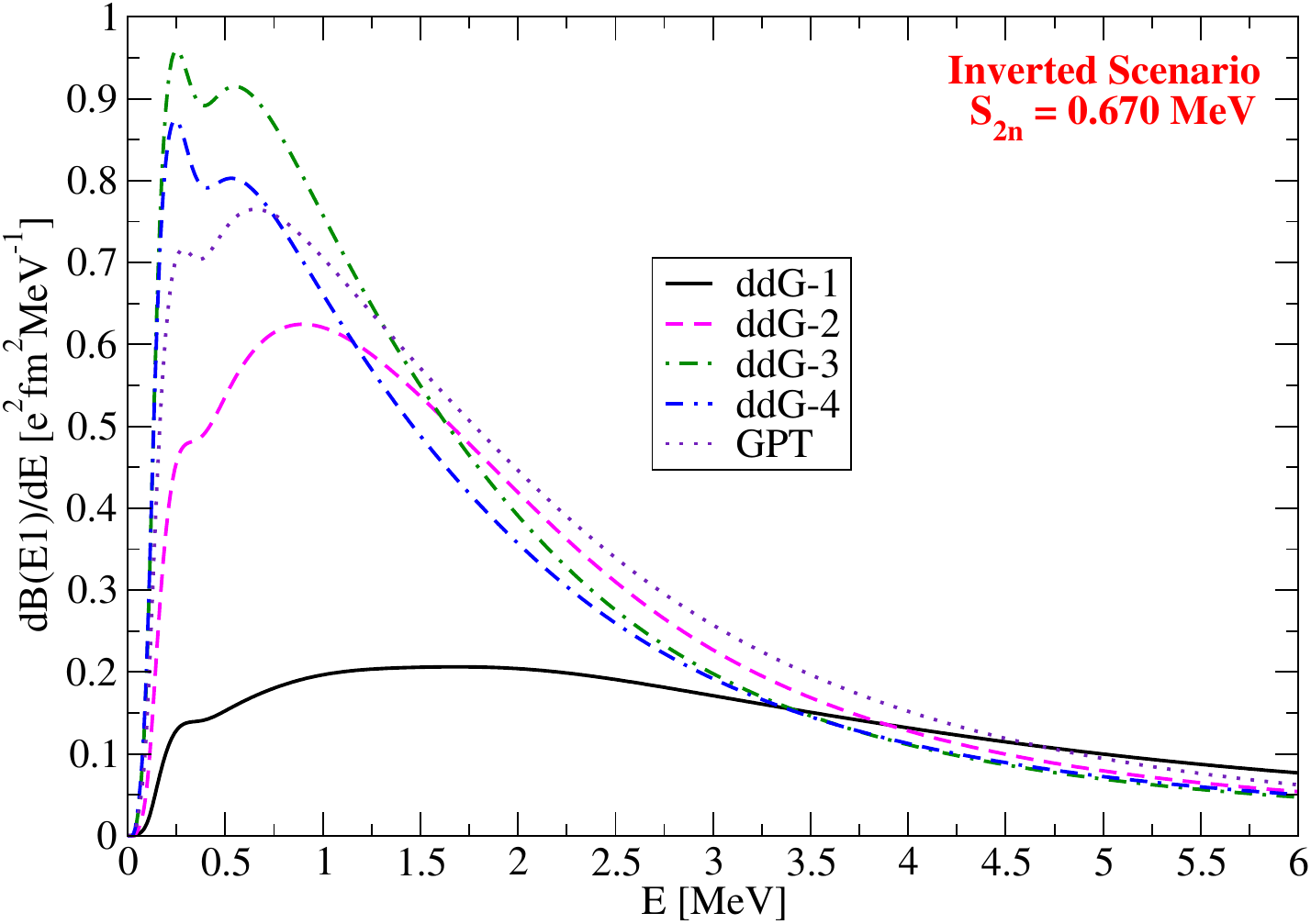}
\caption{$B(E1)$ distribution for $\nucl{40}{Mg}$ for the Inverted scenario using the  central evaluated value for $S_{2n}$.  We use the density-dependent simple Gaussian 1 (solid black line), 2 (dashed magenta line), 3 (dot-dashed green line), and 4 (dot-double-dashed blue line) as discussed in the text, and show the previous GPT plus three-body result as the dotted indigo line.}
\label{fig:BE1ddG}
\end{figure}
Figure \ref{fig:BE1ddG} shows the calculated $B(E1)$ distributions for different neutron-neutron ($n$–$n$) interactions for the Inverted scenario, for  $S_{2n}=0.670$\,MeV. We compare the results obtained using the realistic finite-range GPT interaction supplemented with a three-body force (dotted indigo line) to those from the density-dependent Gaussian potentials defined in Eqs.~\ref{eq:nndensdep} and \ref{eq:ddGpars}: ddG-1 (solid black), ddG-2 (dashed magenta), ddG-3 (dot-dashed green), and ddG-4 (dot-double-dashed blue). 

The $B(E1)$ distribution derived from the GPT interaction corresponds to a ground-state configuration with a $(2p_{3/2})^2$  50.01$\%$ component. In contrast, the ddG interactions yield $(2p_{3/2})^2$ probabilities of 24.93$\%$, 41.20$\%$, 51.15$\%$, and 51.77$\%$ for ddG-1 through ddG-4, respectively. These values indicate that ddG-3 and ddG-4, which employ Gaussian ranges of a few femtometers, closely reproduce the structure obtained with the realistic GPT interaction.

This trend is further reflected in the calculated matter radii: 3.708 fm (ddG-1), 3.757 fm (ddG-2), 3.780 fm (ddG-3), 3.772 fm (ddG-4), and 3.767 fm (GPT). Notably, the shorter-range interactions (ddG-1 and ddG-2, with ranges below 1 fm) result in a reduced $p$-wave and enhanced $f$-wave content in the ground state. This is also reflected in the corresponding $B(E1)$ distributions shown in Figure \ref{fig:BE1ddG}. Analogous to the results presented in Ref.~\cite{Casal19B2020}, the present calculations indicate that the finite range of the $n$-$n$ interaction is a key feature to describe the electric-dipole response of halo nuclei.


\section{Summary}
We investigated the electric-dipole $B(E1)$ response of the neutron-rich nucleus $\nucl{40}{Mg}$ within a three-body 
($\nucl{38}{Mg}+n+n$) hyperspherical framework using an analytical THO basis. We have focused  on the role of 
the $n$-$n$ interaction and the dependence  of the dipole strength on the two-neutron separation energy ($S_{2n}$), which remains
 experimentally uncertain. 
 
 Two types of $n$-$n$ interactions were examined: the finite-range GPT interaction and several density-dependent Gaussian (ddG) central potentials 
 adjusted to reproduce the experimental $n$-$n$ scattering length.
 For all calculations, the $\nucl{38}{Mg}$ was treated as inert and spinless, and the three possible $\nucl{38}{Mg}+n$
scenarios previously identified in Ref.~\cite{SinghPLB2024} were adopted. The strengths of the three-body force or density-dependent term 
were calibrated to reproduce selected values of $S_{2n}$ within the evaluated uncertainty band.

Using the GPT interaction, the calculated $B(E1)$ distributions exhibit strong peaks at low energies, a hallmark of two-neutron halo structure. 
The magnitude of the low-lying dipole strength varies notably among the Normal, Degenerate, and Inverted orbital ordering scenarios. The Inverted case 
shows the largest strength as a result of enhanced $(2p_{3/2})^2$ occupancy and a correspondingly larger matter radius, and again is the most promising scenario for a Borromean halo nucleus. The integrated dipole strengths 
up to 6 MeV exhaust 76–86 percent of the dipole cluster sum rule for these scenarios, reaching values comparable to or exceeding those of established 
two-neutron halo nuclei.

A systematic exploration of $S_{2n}$ confirms the expected correlation between weaker binding, enhanced $p$-wave content, larger matter radius, and stronger dipole response. 
The shallowest case studied ($S_{2n}=0.200$ MeV) yields the most pronounced $B(E1)$ enhancement, exhausting 93 percent of the cluster sum rule below 6 MeV. 
The position and magnitude of the low-energy peak vary smoothly with the assumed ground-state energy, which suggests that the observed structure is not associated with a 
clear dipole resonance. A more stringent resonance identification analysis is needed to settle this point.

A comparison between GPT and ddG interactions in the Inverted scenario highlights the importance of finite-range effects in the $n$-$n$ subsystem. Long-range ddG interactions (ranges of a few fm) reproduce the GPT ground-state structure, matter radius, and $B(E1)$ distribution with high fidelity. Shorter-range versions lead to underestimation of the 
$p$-wave contribution and a noticeable reduction of the dipole strength. These findings reinforce earlier conclusions that a realistic 
finite range in the $n$-$n$ interaction is essential for reliable predictions of halo observables.

Overall, the results show that $\nucl{40}{Mg}$ exhibits characteristics consistent with a two-neutron halo, especially if its actual 
$S_{2n}$ lies near the lower end of the evaluated range. The predicted electric dipole strength provides a clear and measurable signature of this structure; in this case that is a much stronger signal than the matter radius.
Future work should include an explicit treatment of core excitations, improved constraints on $\nucl{38}{Mg}+n$ spectroscopy, and a more rigorous analysis
 of continuum structures to clarify the nature of the observed dipole peak.

\section{Acknowledgments}
 This work was supported by an award from the University of Manchester Faculty of Science and Engineering \textit{Research Dissemination Fund} for research staff (JS), the UK Science and Technology Funding Council [grant number  ST/Y000323/1] (JS and NRW), by the Spanish MICIU/AEI/10.13039/501100011033 and by ERDF/EU [project No.~PID2023-146401NB-I00] (JC), and by the JSPS KAKENHI [Grants Nos.~23K22485, 25K07285, and 25K01005] (WH).


\end{document}